\documentclass[11pt, letterpaper, onecolumn]{article}
\usepackage[utf8]{inputenc}
\usepackage{graphics,graphicx}
\usepackage{amsmath}
\usepackage{amssymb,latexsym,amsopn}
\usepackage[margin=1in]{geometry}
\usepackage{wrapfig}
\usepackage{deluxetable}
\usepackage{times}
\usepackage{xcolor}
\usepackage[colorlinks=true, urlcolor=blue]{hyperref}
\definecolor{dblue}{cmyk}{1, 1, 0.15, 0.17}
\usepackage{sectsty}
\sectionfont{\fontsize{13}{15}\selectfont}
\subsectionfont{\fontsize{12}{15}\selectfont}

\begin{document}

{\fontfamily{ptm}\selectfont
\huge
\begin{center}

 \begin{figure}[ht!]
  \begin{center}
    \includegraphics[width=0.3\textwidth]{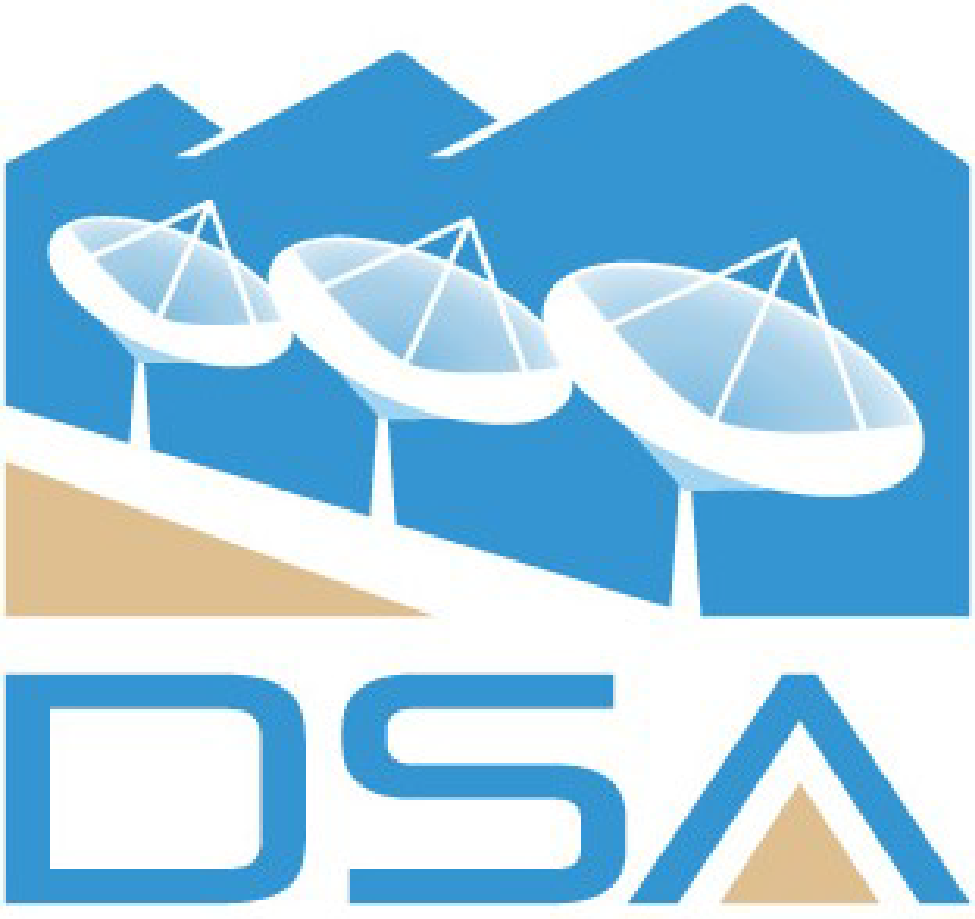}
  \end{center}
  \label{fig:logo}
\end{figure}

Astro2020 APC White Paper \linebreak

The DSA-2000 - A Radio Survey Camera \linebreak
\Large

\noindent \textbf{Type of Activity:} Ground-based project
 
 \begin{figure}[h!]
  \begin{center}
    \includegraphics[width=0.85\textwidth]{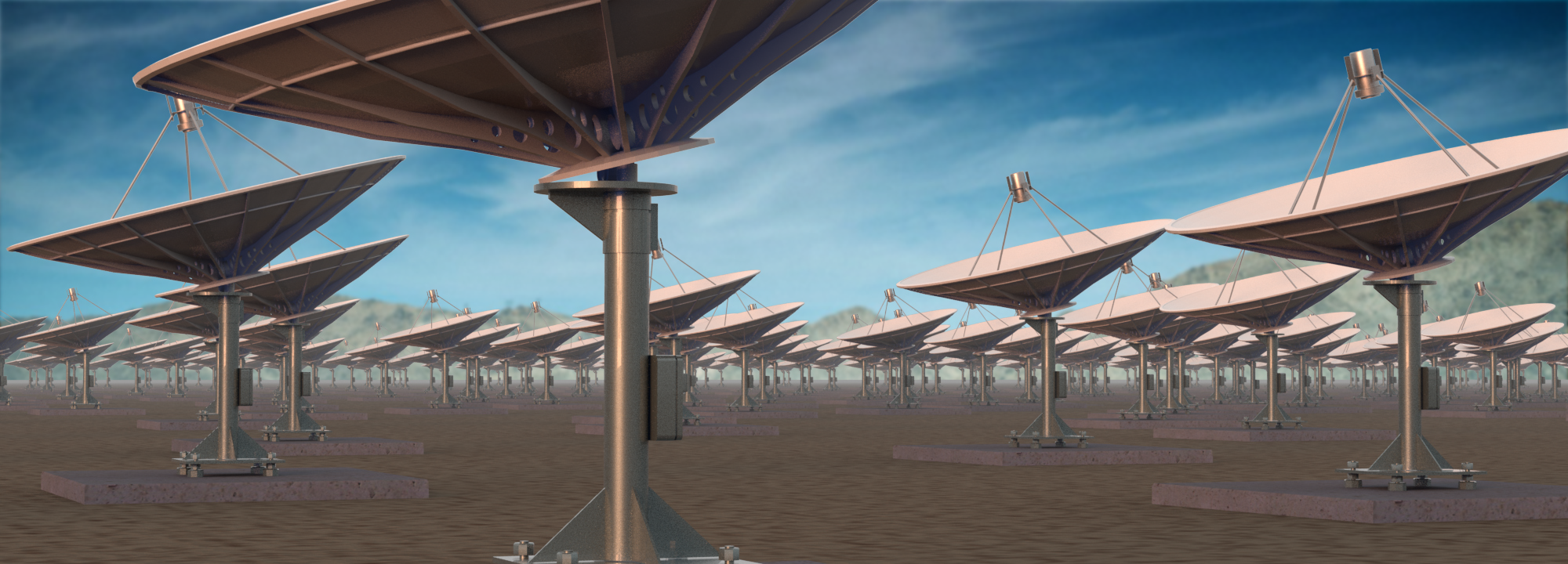}
  \end{center}
  \label{fig:artist}
\end{figure}

\normalsize

  G. Hallinan$^1$, V. Ravi$^1$, S. Weinreb$^1$, J. Kocz$^1$, Y. Huang$^1$, D. P. Woody$^1$, J. Lamb$^1$, L. D'Addario$^1$, M. Catha$^1$, J. Shi$^1$,  C. Law$^1$, S. R. Kulkarni$^1$, E. S. Phinney$^1$, M. W. Eastwood$^1$, K. L. Bouman$^1$ \linebreak
 
 M. A. McLaughlin$^2$, S. M. Ransom$^2$, X. Siemens$^2$, J. M. Cordes$^2$, R. S. Lynch$^2$, D. L. Kaplan$^2$, S. Chatterjee$^2$, J. Lazio$^2$, A. Brazier$^2$ \linebreak
 
S. Bhatnagar$^3$,  S. T. Myers$^3$, F. Walter$^{4,3}$, B. M. Gaensler$^5$
  
  \vspace{0.2cm}
 
 \small 
 
 $^1$\textit{DSA Collaboration} 

$^2$\textit{The NANOGrav Collaboration}

$^3$\textit{NRAO, 1003 Lopezville Road, Socorro, NM 87801, USA}

$^4$\textit{Max Planck Institute for Astronomy, Königstuhl 17, D-69117 Heidelberg, Germany}

$^5$\textit{Dunlap Institute, University of Toronto, 50 St. George Street, Toronto, ON M5S 3H4, Canada}

\end{center}

\normalsize

\textbf{Primary Contacts:} Gregg Hallinan (\href{mailto:gh@astro.caltech.edu}{gh@astro.caltech.edu}), Vikram Ravi (\href{mailto:vikram@caltech.edu}{vikram@caltech.edu})

\pagebreak

\noindent \textbf{Executive Summary:} We present the DSA-2000: a world-leading radio survey telescope and multi-messenger discovery engine for the next decade. The array will be the first true radio camera, outputting science-ready image data over the 0.7--2\,GHz frequency range with a spatial resolution of 3.5\,arcsec. With $2000 \times 5$~m dishes, the DSA-2000 will have an equivalent point-source sensitivity to SKA1-mid, but with ten times the survey speed.  
The DSA-2000 is envisaged as an all-sky survey instrument complementary to the ngVLA, and as a counterpart to the LSST (optical), SPHEREx (near-infrared) and SRG/eROSITA (X-ray) all-sky surveys. Over a five-year prime phase, the DSA-2000 will image the entire sky above declination $-30^{\circ}$ every four months, detecting $> 1$ billion unique radio sources in a combined full-Stokes sky map with 500\,nJy/beam rms noise. This all-sky survey will be complemented by intermediate and deep surveys, as well as spectral and polarization image cubes. The array will be a cornerstone for multi-messenger science, serving as the principal instrument for the US pulsar timing array community, and by searching for radio afterglows of compact object mergers detected by LIGO and Virgo. The array will simultaneously detect and localize $\sim 10^4$ fast radio bursts each year, realizing their ultimate use as a cosmological tool. The DSA-2000 will be proposed to the NSF Mid-Scale Research Infrastructure-2 program with a view to first light in 2026.



\begin{figure}[h]
  \begin{center}
    \includegraphics[width=0.95\textwidth]{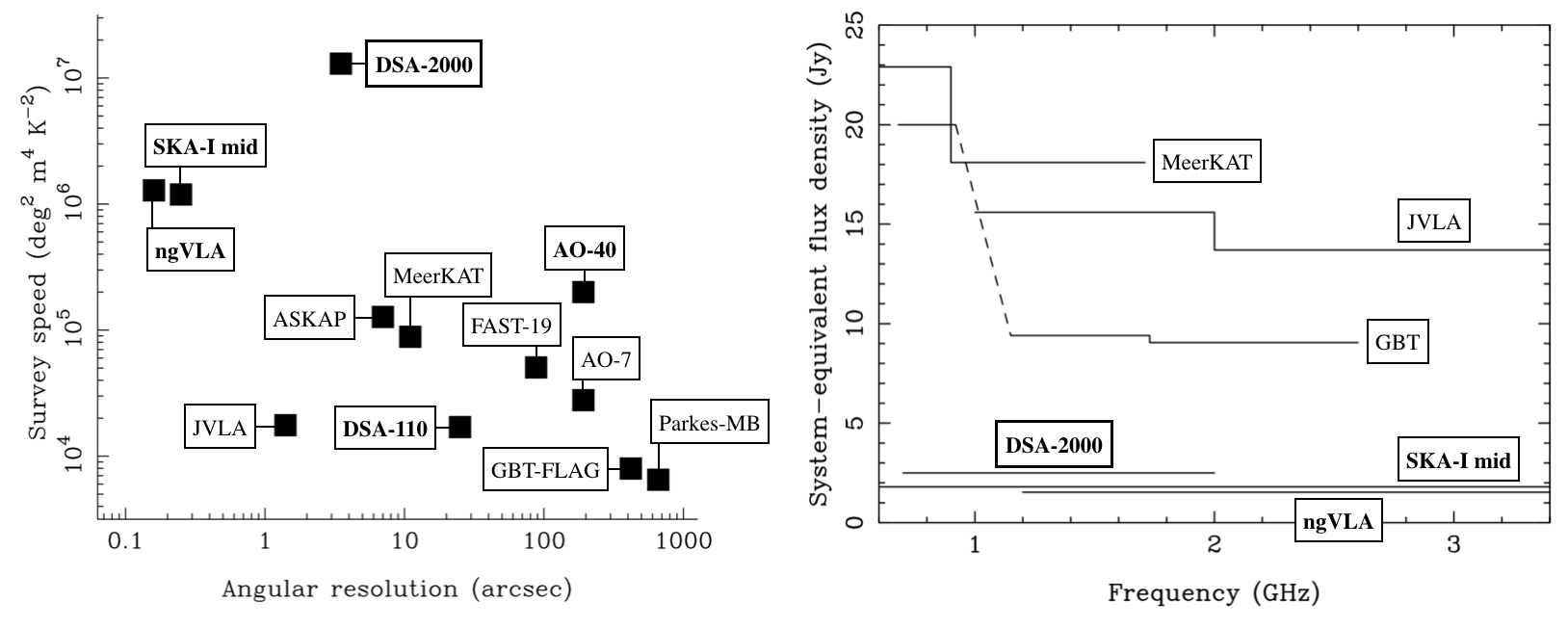}
  \end{center}
  \caption{\color{dblue} \small {\em Left:} The survey speed and angular resolution of the DSA-2000 compared with other operational (normal text) or upcoming (bold font) telescopes operating at 1.4\,GHz. {\em Right:} The system-equivalent flux density (SEFD; inversely proportional to sensitivity) of the DSA-2000 and other steerable radio telescopes. Sensitivities are shown in the specific frequency ranges accessed by each telescope. 
  }
  \label{fig:sens}
\end{figure}

\vspace{-0.6cm}
\section{Key Science Goals and Objectives}


Surveys with highly optimized, dedicated telescope instrumentation, and with a focus on delivering science-ready data products to the community, are a mainstay of modern astronomy. They simultaneously democratize and increase the diversity of groundbreaking research by offering easier access to the most important data sets. 
The US has led the world in delivering optimized surveys with revolutionary data accessibility that continue to transform our perspective on the Universe: SDSS, ZTF, PanSTARRS, and the Dark Energy Survey in the optical; UV surveys with GALEX; and a $\gamma$-ray survey with {\em Fermi}. A highly successful survey of the gravitational-wave (GW) sky has commenced with the aLIGO detectors. The LSST was the highest priority recommendation of the 2000 and 2010 Astronomy and Astrophysics Decadal Surveys, and will commence operations in 2023. 
The LSST, together with surveys by ZTF, DESI, WFIRST, and SPHEREx, will see US leadership in optical and infrared surveys continue into the 2020s.


US leadership in surveys extends into the radio band, most notably with the VLA. The 1.4\,GHz NRAO VLA Sky Survey (NVSS, early 1990s) cataloged $\sim2$ million sources over 82\% of the sky with moderate angular resolution  (45\,arcsec) (Condon et al. 1998). The Faint Images of the Radio Sky at Twenty-Centimeters (FIRST) survey (VLA, 1.4\,GHz, 1990s) observed the same $\sim 10,000$\,deg$^2$ region of the sky as SDSS with 5\,arcsec angular resolution, better suited for matching with optical survey data (Becker et al. 1995). 
The ongoing VLA Sky Survey (VLASS) will conduct a three-epoch survey of the sky north of $-40^{\circ}$ between $2-4$\,GHz with 2.5\,arcsec angular resolution. VLASS provides full polarization data products, and its cadenced nature enables unprecedented sensitivity to radio transients (Mooley et al. 2016, Law et al. 2018). The scientific legacy of NVSS/FIRST has been immense and grows rapidly today; the two survey publications are the most cited papers produced by the VLA (Becker et al. 1995; Condon et al. 1998). \textit{This is despite the fact that the VLA is poorly optimized for surveys (Figure~\ref{fig:sens})}. 


{\bf We present the Deep Synoptic Array 2000-antenna concept (DSA-2000): a world-leading radio survey camera for the 2020s.} The array will consist of $2000 \times 5$~m steerable dishes instantaneously covering the 0.7–2\,GHz frequency range (Table~\ref{tab:t1}). 
Designed from the ground up for survey science, the DSA-2000 delivers unparalleled survey speed via a low-cost platform, with $10 \times$ the survey speed of the SKA-I mid-frequency array (Figure~\ref{fig:sens}). 


\begin{figure}[h]
\centering
\includegraphics[width=0.9\textwidth]{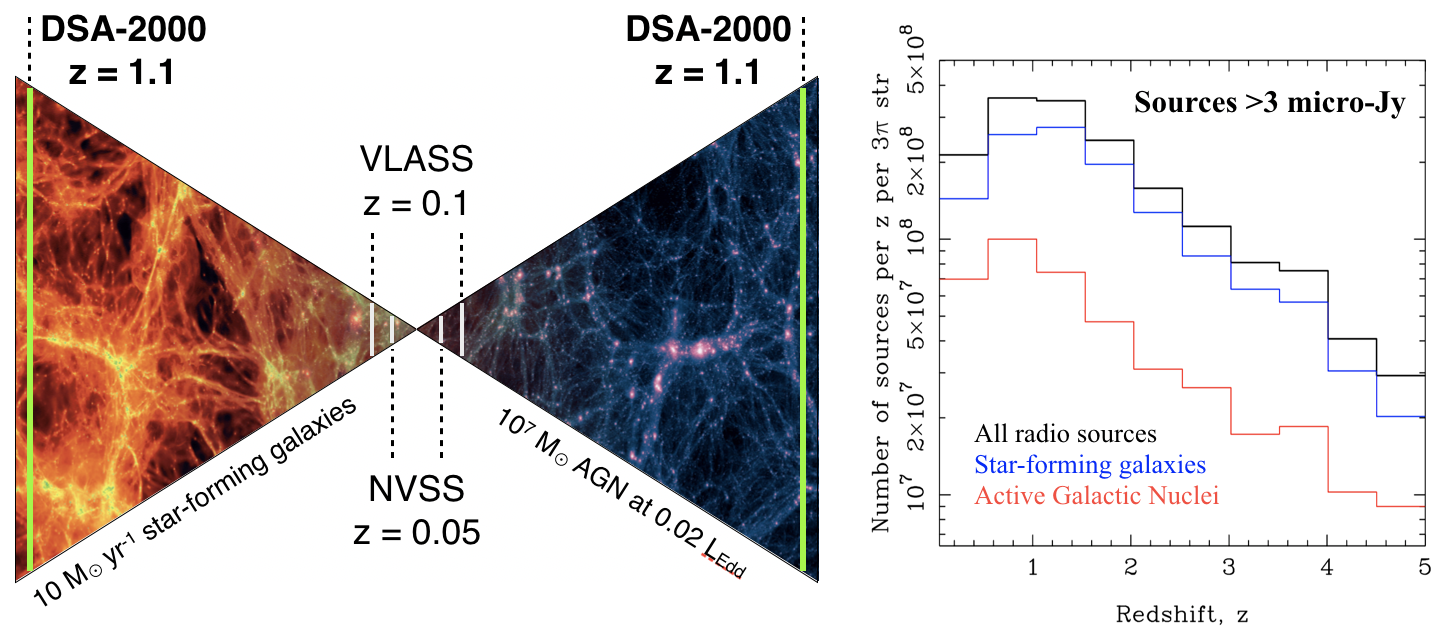}
\caption{\color{dblue} \small {\em Left:} Detection horizons of the NVSS, VLASS and DSA-2000 surveys to star-forming galaxies and low-luminosity AGN. We consider the typical radio luminosities of 10\,$M_{\odot}$\,yr$^{-1}$ galaxies on the left, and $10^{7}M_{\odot}$ AGN accreting at $0.02L_{\rm Edd}$ on the right. The DSA-2000 Cadenced All-Sky Survey, when complete, will detect these systems to redshifts $z=1.1$, probing a $2500\times$ larger volume than the final VLASS data set. Background image credit: Illustris Simulation. {\em Right:} Predicted redshift distribution of sources ($>3\,\mu$Jy) in the DSA-2000 Cadenced All-Sky Survey stack (Wilman et al. 2008). The billion sources detected by DSA-2000 will be dominated by star-forming galaxies (blue) at all redshifts.} 
\label{fig:sf_agn}
\end{figure}

\noindent {\bf KEY OBJECTIVE 1: The Cadenced All-Sky Survey ($\sim65\%$ of time).} The DSA-2000 will  survey the entire sky north of $-30^{\circ}$. Image mosaics for $3\pi$\,sr will be produced for each of 16 epochs spaced by 4 months, using $\sim 6,000$ 15-min pointings. Each epoch will be archived as images in $10 \times$ 128\,MHz bands, as well as a single wideband image. The latter will be a trillion pixel map with rms noise of $2\,\mu$Jy/beam.  The final stacked all-sky image will have an rms noise of 500\,nJy/beam, and will contain $\sim 10^9$ discrete radio sources (Figure~\ref{fig:sf_agn}). The map will be served to the community as a fully calibrated data set accompanied by an associated source catalog with a 10-point SED and 16-point light curve for each object.



The Cadenced All-Sky Survey will also produce a spectral image cube with $8,000 \times 162.5$\,kHz channels across the entire 0.7--2\,GHz band. Windows of higher resolution (4,000 channels at 10\,kHz; 2\,km/s) data around the HI and OH lines will trace ISM in our Galaxy and in galaxies out to 100\,Mpc. A full polarization (Stokes IQUV) combined full-band data cube ($1000 \times 1.3$~MHz channels) will also be produced. These data are too large for a public archive, but will be made available to the community for source finding and associated analysis.

{\it The DSA-2000 will increase the known sample of radio sources by a factor of 1000.} The survey will enable the most complete census yet of active galaxies (Maglicchetti et al. 2018, Mezcua et al. 2019), including low-luminosity AGN at the low end of the supermassive black hole mass function, probing seed mechanisms (Astro2020 SWPs \href{http://surveygizmoresponseuploads.s3.amazonaws.com/fileuploads/623127/4458621/64-143cbe52c445b35eb6fe3f82ee1c0061_IMBH_Astro2020submit.pdf}{Greene et al.}, \href{http://surveygizmoresponseuploads.s3.amazonaws.com/fileuploads/623127/4458621/198-4ca093aadad8099fb10d7d7cdec79058_NatarajanPriyamvada.pdf}{Natarajan et al.}). The stacked continuum images will provide an unobscured census of cosmic star-formation (e.g., Cotton et al. 2018), detecting galaxies forming stars at a few $\times100\,M_{\odot}$\,yr$^{-1}$ at redshifts $z\sim6$ (Decarli et al. 2017). 
High-spectral resolution data will enable resolved HI and OH kinematics within 100\,Mpc, tracing the role of gas inflow in star formation on all scales. The lower resolution data will enable a catalog of HI galaxies at $z<1$, with applications ranging from the integrated Sachs-Wolfe Effect to providing a reference catalog for localization of nearby GW events. 
{\em The cadenced nature of the DSA-2000 continuum survey makes it a true synoptic counterpart to the LSST.}  The individual DSA-2000 survey epochs will detect slowly varying radio transients in a 1700 times larger volume than VLASS and over five times the number of epochs. These data will also provide a new window on stellar activity and compact object physics in our own Galaxy. The DSA-2000 data will also be a rich resource for the exploration of the unknown (Astro2020 SWP \href{http://surveygizmoresponseuploads.s3.amazonaws.com/fileuploads/623127/4458621/64-6adac0450e8fa3f3cf11feeaf254dd6d_SiemiginowskaAneta.pdf}{Siemiginowska et al.}). 


\vspace{6pt}
\noindent {\bf KEY OBJECTIVE 2: LSST Deep Drilling Fields  ($\sim5\%$ of time).} The DSA-2000 will observe three LSST deep drilling fields (XMM-LSS, Extended Chandra Deep Field-South, COSMOS) using single pointings at a daily cadence. Each of the final $\sim 10$\,deg$^2$ images will be confusion-noise limited at the $\sim 100$\,nJy/beam level (Condon et al. 2012). Together with  multi-band LSST photometry and existing panchromatic data, the DSA-2000 deep images will provide sensitivity to star-formation rates of $\sim100 M_{\odot}$\,yr$^{-1}$ at $z=6$, and rates of $\sim10 M_{\odot}$\,yr$^{-1}$ at $z=2$. An additional 2,000\,deg$^2$ will be available with typical rms $\sim 200$\,nJy/beam  due to frequent observations of $\sim 200$ pulsar timing fields, providing an intermediate tier between all-sky and deep fields.

\vspace{6pt}
\noindent {\bf KEY OBJECTIVE 3: The Pulsar Timing Array ($\sim25\%$ of time).} The DSA-2000 will be the principal instrument for the US-led NANOGrav pulsar timing array, with pulsar timing making up approximately 25\% of the prime phase of the project (Figure~\ref{fig:fig_mm}, Astro2020 SWPs \href{http://surveygizmoresponseuploads.s3.amazonaws.com/fileuploads/623127/4458621/28-83ef6f2f893ed0a5c0b90ee47c26bbb5_ColpiMonica.pdf}{Colpi et al.}, 
\href{http://surveygizmoresponseuploads.s3.amazonaws.com/fileuploads/623127/4458621/37-5fe3ca0038a60c3bfdf681060e5a47b8_CordesJamesM.pdf}{Cordes et al.}, 
\href{http://surveygizmoresponseuploads.s3.amazonaws.com/fileuploads/623127/4458621/100-e33b2e01346bb624397d65e8cf8adcb8_FonsecaEmmanuel.pdf}{Fonseca et al.},
\href{http://surveygizmoresponseuploads.s3.amazonaws.com/fileuploads/623127/4458621/188-40aea7e2f5f24efd72a7a8c276b177b9_KelleyLukeZ.pdf}{Kelley et al.},  \href{http://surveygizmoresponseuploads.s3.amazonaws.com/fileuploads/623127/4458621/137-64740a44cec45a09a10c74149218dd8d_SiemensXavier.pdf}{Siemens et al.}, 
\href{http://surveygizmoresponseuploads.s3.amazonaws.com/fileuploads/623127/4458621/137-eeea74d107b9921cb5788e3f47cf1dd6_TaylorStephenR.pdf}{Taylor et al.}). By the mid-2020s, NANOGrav will likely have detected the stochastic GW background from binary supermassive black holes. Observations at weekly to monthly cadence of a large number of millisecond pulsars (MSPs) will be necessary to measure its spectrum, search for anisotropies, and test for deviations from the predictions of general relativity. NANOGrav will also be in the regime where a detection of a single, continuous wave source is likely. Observations at higher (optimally daily) cadences of fewer (perhaps dozens) MSPs will be necessary to enable the transformative multi-messenger astrophysics that will result from single-source detection. 
The 0.7--2\,GHz frequency band is the optimal range established through simulations accounting for pulsar spectra, timing precision, ISM effects, and various sources of noise. 

\begin{figure}[h]
	\centering
	\includegraphics[width=0.95\textwidth]{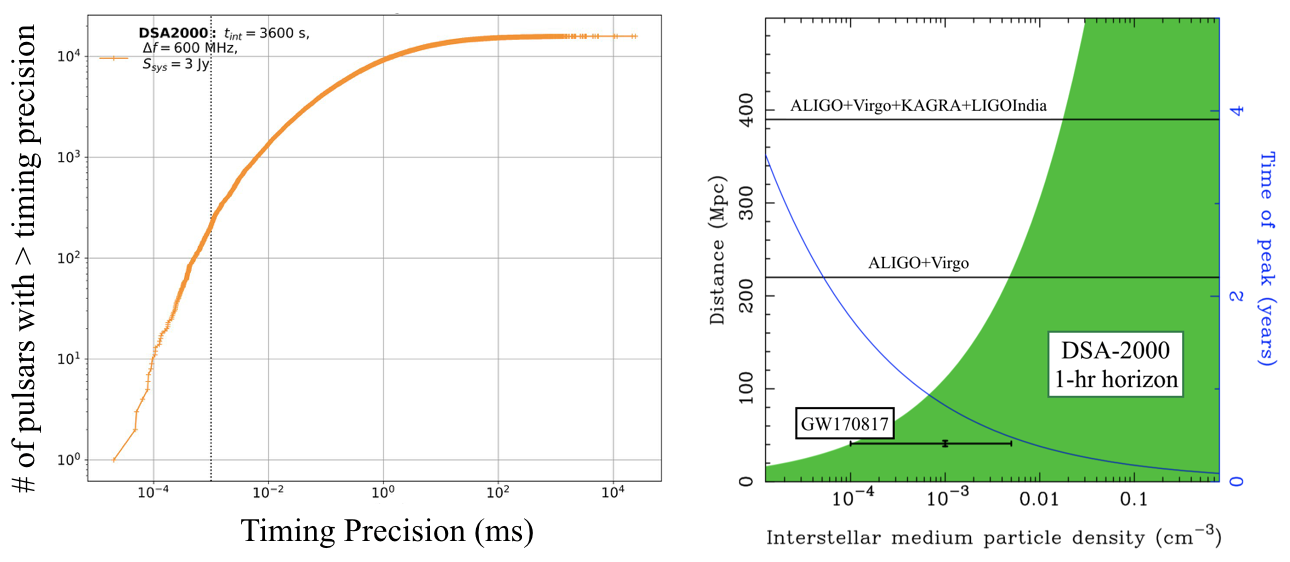} 
	\caption{\color{dblue} \small {\em Left:} The number of MSPs in $3\pi$ str of the sky above given timing precisions, observed with the DSA-2000 in 3600\,s (credit: the NANOGrav Collaboration). {\em Right:} The detection horizon (left black axis; green shading) of the DSA-2000 to radio emission associated with mildly relativistic ejecta from binary neutron star mergers, for different ISM densities (Nakar \& Piran 2011). No relativistic jet is included; this would significantly increase the horizon. The GW detection horizons of the current aLIGO / Virgo and the expected five-detector configuration in the 2020s are shown as horizontal lines. The inferred ISM density surrounding the binary neutron star merger GW\,170817 is shown as an error range (Mooley et al. 2018), and the blue trace / right blue axis shows the timescale of the radio transient emission.}
	\label{fig:fig_mm}
\end{figure}

\noindent {\bf KEY OBJECTIVE 4: Compact-Object Merger Follow-up ($\sim5\%$ of time).} 1\,hr per day will dedicated to discovering and monitoring the afterglows of GW events detected by aLIGO / Virgo, and eventually a five-detector GW observatory (Astro2020 SWP \href{http://surveygizmoresponseuploads.s3.amazonaws.com/fileuploads/623127/4458621/100-cfd8583e04811d53e880c0c4e04d2fc7_CorsiAlessandra.pdf}{Corsi et al.}). In the latter case, the median $90\%$ credible localization region is 9--12\,deg$^2$. The DSA-2000 is ideally specified to follow-up such events, imaging an entire localization region to $\sim 1\,\mu$Jy/beam rms in 1\,hr.   The radio observations of the neutron-star merger GW\,170817 were crucial in identifying the energetics and geometry of the ejected material (Hallinan et al. 2017, Mooley et al. 2018). This in turn enabled constraints on the orientation of the binary, significantly reducing uncertainties in using the event as a `standard siren' for cosmological measurements (Hotokezaka et al. 2019). The DSA-2000 will enable similar scientific outcomes for several GW-detected neutron-star mergers in the future, without an initial detection at other wavelengths. This is particularly important for neutron star - black hole mergers (Nakar \& Piran 2011). 

\vspace{6pt}
\noindent {\bf KEY OBJECTIVE 5: Fast Radio Bursts (FRBs) and VLBI.} A subset of the DSA-2000 will commensally search the entire 10.6\,deg$^2$ field of view for short duration transients, particularly FRBs and single pulses from pulsars. The DSA-2000 will detect and localize FRBs at a rate of $\sim 10^4$/year, primarily for the characterization of the circum-/intergalactic medium, for an investigation of compact object (e.g., primordial black hole) models for dark matter, and as a cosmological probe (Astro2020 SWP \href{http://surveygizmoresponseuploads.s3.amazonaws.com/fileuploads/623127/4458621/28-adf50e8eab168a4d02336a4a568b47b1_RaviVikram.pdf}{Ravi et al.}). FRBs may also trace the most extreme magnetars in their nascent stages (Astro2020 SWP \href{http://surveygizmoresponseuploads.s3.amazonaws.com/fileuploads/623127/4458621/132-82bf097f4299e0fe04749740236ef9af_LawCaseyJ.pdf}{Law et al.}). The DSA-2000 beamformer will commensally provide the capability to conduct VLBI observations in combination with pre-existing VLBI infrastructure, such as the VLBA, or with a partner array of DSA antennas installed at US university sites. This will be used to provide VLBI observations of GW events, and MSPs for distance measurements towards characterizing GWs from individual binary supermassive black holes. 

\begin{figure}[htbp]
	\centering
	\includegraphics[width=0.95\textwidth]{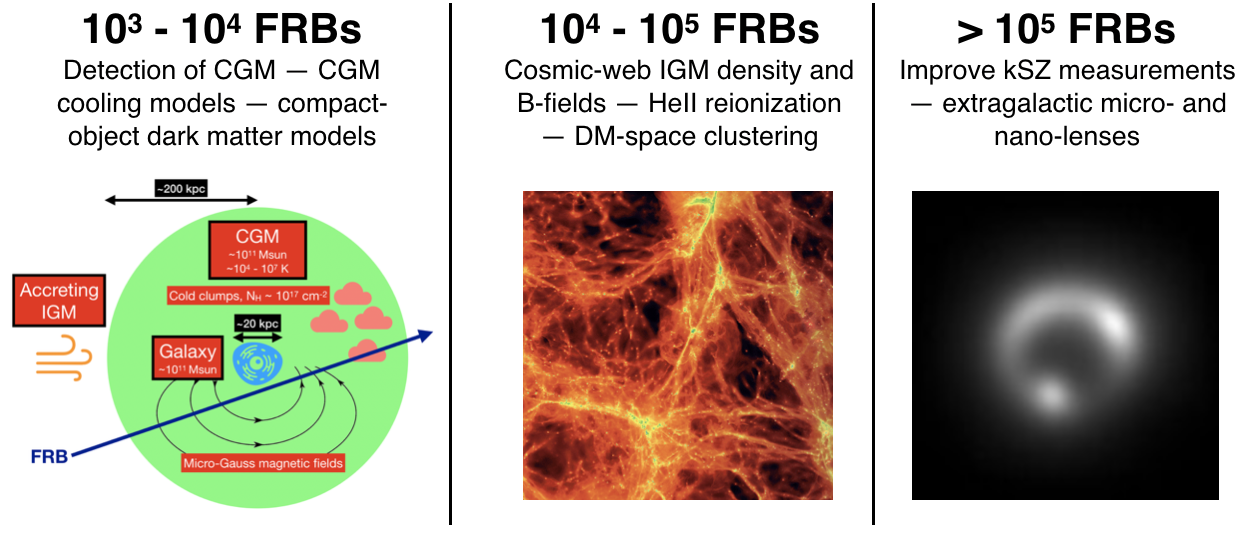} 
	\caption{\color{dblue} \small Summary of the scientific potential of large numbers of interferometrically localized FRBs. The DSA-2000 is expected to detect and localize $\sim10^{4}$ FRBs per year of operations. Figure credits: Illustris Simulation.}
	\label{fig:frb}
\end{figure}

\section{Technical Overview}

\subsection{DSA-10 and DSA-110} The DSA-2000 will be the third iteration of the DSA concept, consisting of low-cost antenna/receiver packages linked by an RF over fiber (RFOF) network to a powerful digital backend. The concept builds on progress with the DSA-10 (operational) and DSA-110 (under construction). The DSA-10 (Kocz et al. 2019) uses 10 off-the-shelf 4.5-m dishes and mounts, with manually adjusted pointing (Figure~\ref{fig:overview}), and custom feeds and commercial ambient-temperature low-noise amplifiers (LNAs) resulting in an overall system temperature of $\sim60$~K and aperture efficiency of $62\%$. The DSA-10 antenna and receiver package was built for a unit cost of $<\$4000$, including labor, with a comparable per-antenna digital backend cost. The DSA-10 has been searching for FRBs at frequencies of $1.28 - 1.53$\,GHz  since June 2017, {\em and recently successfully localized an FRB to its host galaxy (Ravi et al. 2019).}


The fully funded (NSF MSIP) DSA-110 is a significant evolution from the DSA-10. Consisting of $110 \times 4.75$\,m dishes distributed across 2.5\,km, each antenna has a motorized elevation drive and a custom-designed LNA (noise temperature of 11\,K; Figure~\ref{fig:receiver}). The final system temperature will be $< 30$\,K (verified by current on-sky tests), with aperture efficiency of $70\%$ operating in the same $1.28 - 1.53$\,GHz band as DSA-10. The DSA-110 dish, mount, motorized elevation drive and receiver have a unit cost of $<\$7500$, including labor. Again, a similar digital-backend cost per antenna is anticipated. The project passed Preliminary Design Review (PDR) in January 2019. The DSA-110 will deliver a sample of $>300$ FRBs localized to $< 3-$arcsec within a 3-yr science program, enabling transformational advances in our understanding of the origins of FRBs, and of the unseen baryons in the circum-/intergalactic medium. FRB alerts will be issued via the VOEvent service to the entire astronomical community to ensure rapid follow-up of localized FRBs.

\vspace{6pt}
\subsection{DSA-2000} A few key technological advances over the DSA-110 are required for the DSA-2000. The antennas will be 5\,m dishes serviced by fully motorized elevation, azimuth and tracking capable mounts / drives. The receiver will incorporate a wider band feed and ambient temperature LNA. The baseline specification assumes a system temperature of 25~K, with aperture efficiency of $70\%$ (SEFD 2.5~Jy; Table~\ref{tab:t1}).
The digital backend will support three simultaneous modes: i) a 4096-input full cross-correlator and imager; ii) a beamformer for pulsar timing; and iii)  a partial-array multiple-beam system for FRB searches. An existing conceptual design is in place, making use of 2019 hardware and utilizing a similar ``FX" architecture to that chosen for the DSA-110. 

\begin{figure}[h]
  \begin{center}
    \includegraphics[width=0.95\textwidth]{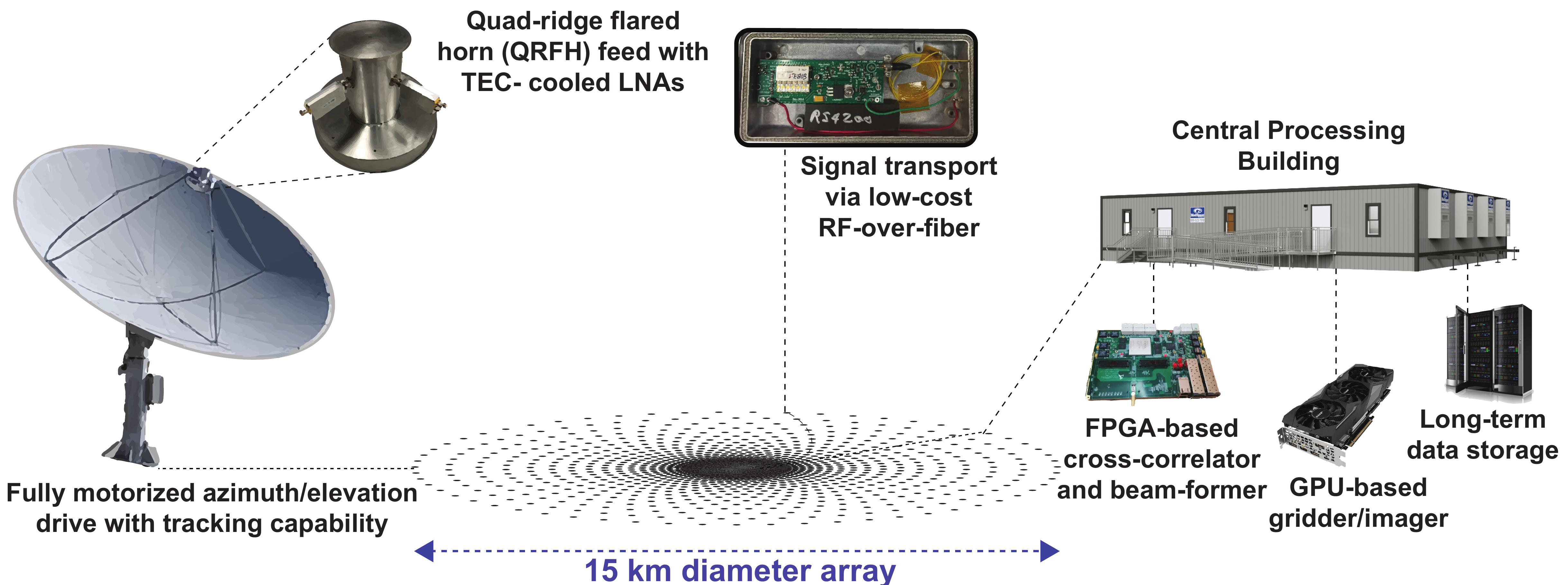}
  \end{center}
  \caption{\color{dblue} \small The DSA-2000 will consist of $2000 \times 5$\,m fully automated, steerable dishes distributed across 15\,km in a configuration optimized to minimize sidelobes in 15-min integrations. Each antenna is serviced by a quad-ridge horn feed and an ambient temperature LNA. Low cost RFOF modules and a distributed network of optical fiber will transport DSA-2000 data to the Central Processing Building, housing the cross-correlator, beam-former and imaging backend.}
  \label{fig:overview}
\end{figure}

\begin{table}[h]
\caption{\small Baseline DSA-2000 specifications. Assumes $65\%$ usable bandwidth and 20\% time overheads.}
\normalsize
\begin{center}
 \begin{tabular}{||c| c||} 
 \hline
Quantity & Value \\
\hline\hline
Reflectors & $2000 \times 5$-m dishes \\
Frequency Coverage & 0.7--2\,GHz \\
Bandwidth & $1.3$\,GHz \\
Field of View & 10.6\,deg$^2$ \\
Spatial Resolution & 3.5\,arcsec \\
System Temperature & 25\,K  \\
Aperture Efficiency & $70\%$ \\
System-Equivalent Flux Density (SEFD) & 2.5~\,Jy \\
Survey Speed Figure of Merit & $1.3 \times 10^7$\,deg$^2$\,m$^4$\,K$^{-2}$\\
Continuum Sensitivity (1 hour) & $1\,\mu$Jy\\
All-Sky Survey (per epoch) & $30,000$\,deg$^2$ @ $2\,\mu$Jy/bm \\
All-Sky Survey (combined) & $30,000$\,deg$^2$ @ 500\,nJy/beam \\
Pulsar Timing Fields (Intermediate) & 2000~deg$^2$ @ 200\,nJy/beam \\
Deep Drilling Fields & $30$\,deg$^2$ @ 100\,nJy/beam \\
Brightness Temperature ($1\sigma$)  & 5\,mK \\
Number of Unique Sources  & $>1$~billion \\
\hline
\end{tabular}
\end{center}
\label{tab:t1}
\end{table}

{\em The DSA-2000 will serve the community with images, not interferometric visibilities, making its data products more easily usable by a wider user base than for current or planned radio interferometers.}
Unlike existing radio telescopes, data editing, calibration and gridding/imaging of the DSA-2000 will be applied in real-time on a GPU cluster integrated into the digital backend.
This is enabled by the dense array configuration, optimized for low synthesized beam sidelobes in 15-minute integrations. This gives a dynamic range of $> 10^6$ (likely $> 10^7$ for the final array configuration) between the brightest source and the sidelobes in the $\sim10$~deg$^2$ images of the DSA-2000 (Figure~\ref{fig:config_psf}). This removes the need for visibility-based blind deconvolution, the  computationally expensive mainstay of radio interferometric imaging. In the absence of non-linear deconvolution, calibration/imaging is completely deterministic and highly optimized on GPU platforms.  The residual sidelobes from bright sources will dominate the noise for a few-\% of survey data. These data can be recovered through computationally cheap image-plane deconvolution on the final data products (Figure~\ref{fig:config_psf}). For the 40 brightest sources in the sky ($>10$\,Jy/beam), externally-derived source models can be subtracted from the visibilities prior to gridding, if necessary. 

\begin{figure}[t]
	\centering
	\includegraphics[width=\textwidth]{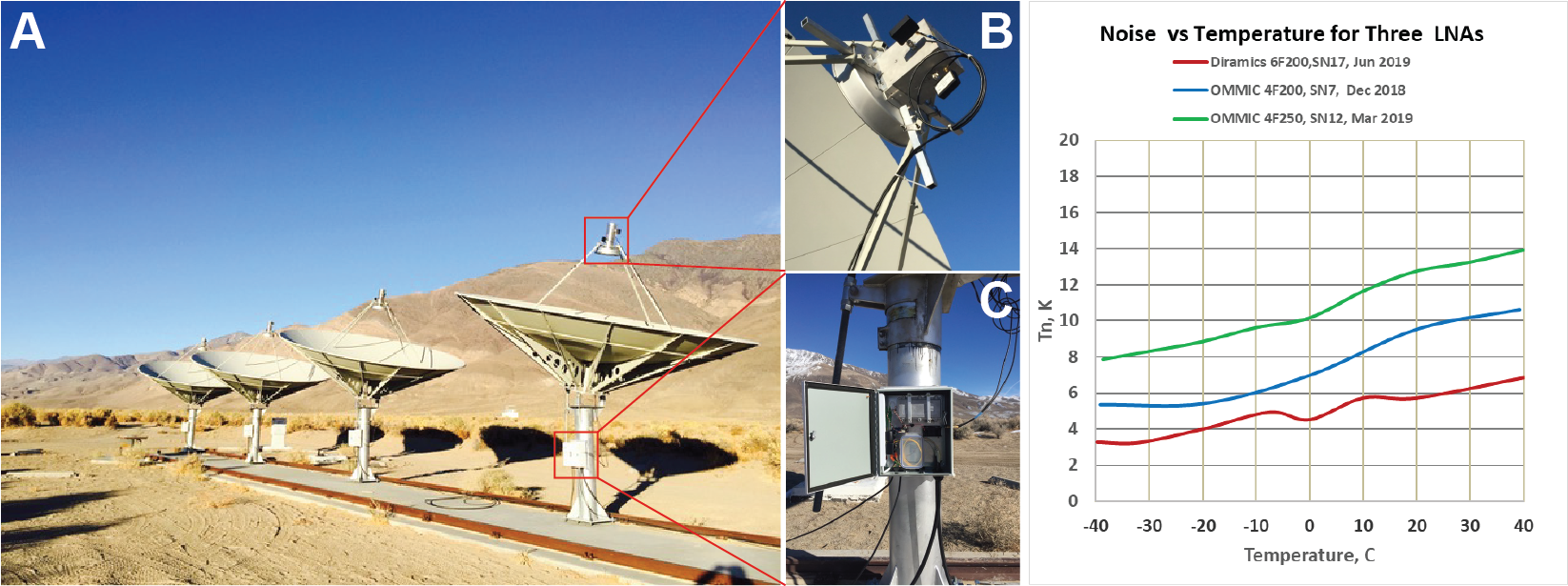} 
	\caption{\color{dblue} \small \textbf{A:} Four of the antennas of the DSA-10 located at the Owens Valley Radio Observatory (OVRO). \textbf{B:} The custom-designed feed attached to an antenna of the DSA-10. \textbf{C:} The front-end enclosure attached to one of the DSA-10 antennas, which houses the front-end electronics. {\em Right:} Noise temperature as a function of temperature for LNAs tested during the DSA-110 design phase. An LNA with a Diramics transistor has been tested with 6~K noise temperature at room temperature and will serve as the platform for the DSA-2000 wideband LNA design.}
	\label{fig:receiver}
\end{figure}

The DSA-2000 array requires a flat plain spanning approx. 15\,km, in an RFI-quiet environment that can be accessed with sufficient power and data transport infrastructure.  A number of potential sites have been identified in both California and Nevada, typically in valleys along the Sierra, Inyo and White mountain ranges. The plains of San Agustin in New Mexico, home to the VLA, is a possible alternative site with existing infrastructure. 

\begin{figure}[ht!]
	\centering
	\includegraphics[width=\textwidth]{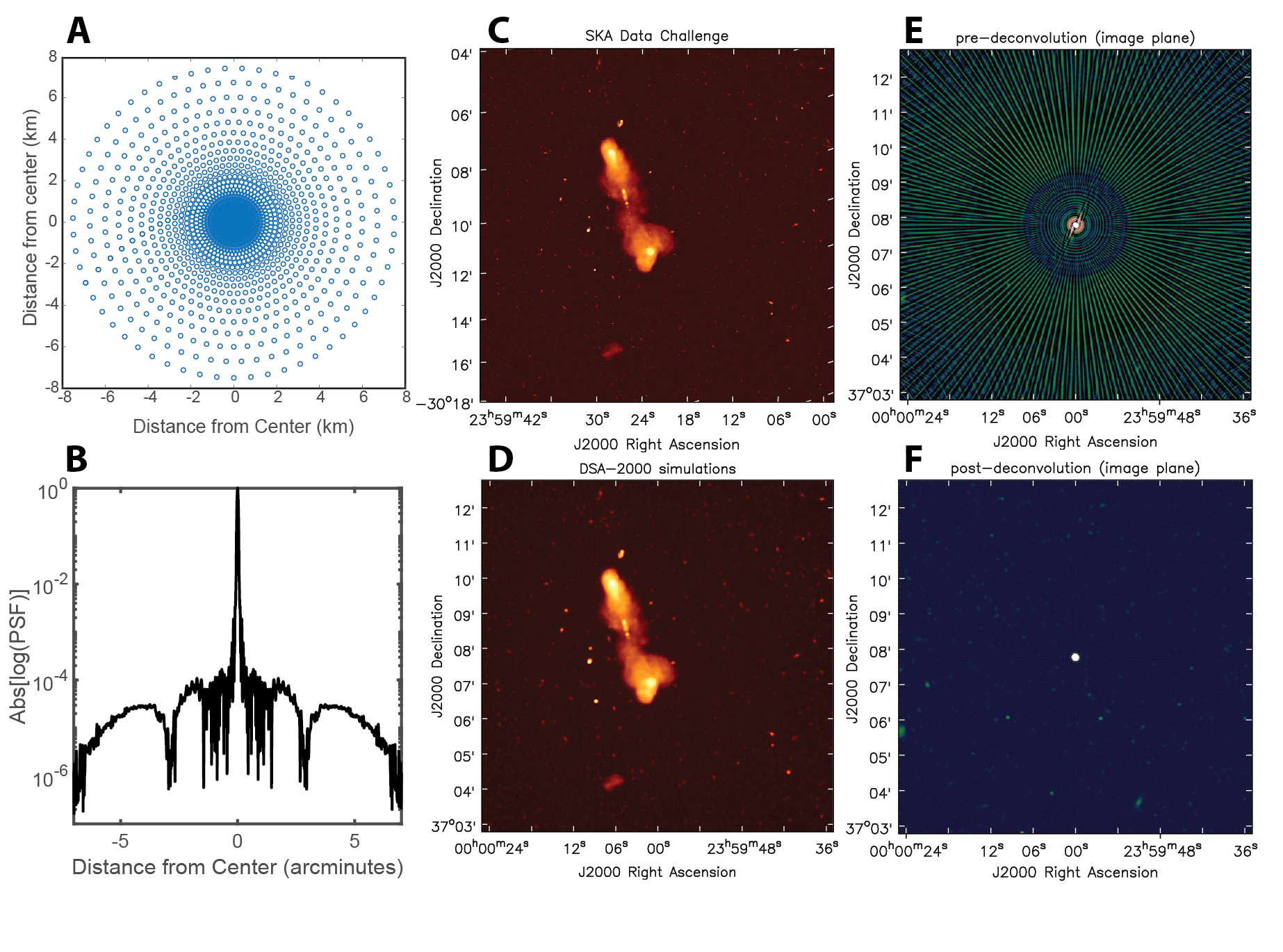} 
	\caption{\color{dblue} \small{\em A:} A preliminary configuration for the DSA-2000 spanning a 15\,km diameter area. The final configuration will be optimized for 15-min tracks. {\em B:} A cut through the point spread function for a 15-min track. {\em C:} A sub-image from the SKA Data Challenge adjusted to the DSA-2000 frequency band. {\em D:} An image created from a simulation of a 10 second snapshot with DSA-2000 using The SKA Data Challenge image as the model sky, to demonstrate the capability of DSA-2000 for deconvolution-free imaging. Gaussian noise of 2$\mu$Jy has been added to both the original and convolved image to represent the expected noise in a 15-minute track with DSA-2000. All sources from the original image are detected in the DSA-2000 image. {\em E:} An image produced from a simulation of the same data at the location of the brightest source in the SKA Data Challenge, scaled to DSA frequencies ($\sim 320$\,mJy at 1.35 GHz) and created without visibility-based deconvolution. {\em F:} The simulated DSA-2000 image of the bright source after 5,000 iterations of image plane deconvolution ($<10$\,s on a single core, 64\,GB of RAM) removes all sidelobes with a resulting final dynamic range of $5 \times 10^5$. Note this is for a 10-second snapshot with DSA-2000. The sidelobes for a 15 minute track will be 10x lower enabling image-plane deconvolution for sources as bright as 10~Jy.}
	\label{fig:config_psf}
\end{figure}


The DSA-2000 will deliver a public archive of the data drawn from the Cadenced All-Sky Survey including a combined trillion-pixel ($8.2\times10^{11}$ pixels, $\sim3.3$\,TB) continuum image of $3\pi$ sr of the sky, and an automatically extracted catalog of of $\sim10^{9}$ sources with flux densities in ten frequency bands over 16 epochs.  Funds will be requested to serve these data to the community in an easily accessible and science-ready manner, via a dedicated user interface. Additional science datasets include commensal FRB observations, GW follow-up data, deep and intermediate field data, $4000\times10$-kHz channels capturing the HI and OH lines in the local volume ($\lesssim100$\,Mpc), $8000 \times 162.5$\,kHz channels for higher-redshift spectral line studies and polarization data ($1000 \times 1.3$\,MHz channels) for a total data volume of $\sim70$\,PB). The DSA-2000 pulsar timing datasets will be processed and served by the NANOGrav collaboration.

\section{Technology Drivers}

A number of technology drivers will be prototyped during the DSA-2000 design phase: 

\begin{itemize}

\item The DSA-2000 requires a robust low-cost azimuth/elevation and tracking drive, building on the elevation drive of the DSA-110.

\item Requirements for dish surface accuracy, beam uniformity, uniformity between dishes, dish stiffness (gravitational deformity) and thermal stability are much more stringent than for the DSA-110, and will be defined based on required a-priori beam knowledge to deliver the dynamic range ($>10^5$) in \textit{direction-dependent} calibration of the DSA-2000. We note that the DSA-110 dish greatly exceeds design requirements and will likely meet many of the specifications for DSA-2000. 

\item An additional requirement of the DSA-2000 is to ensure residual calibration errors (both phase and amplitude) are sufficiently small to ensure that errors in the image domain are below the synthesized beam sidelobes. For example, the average residual phase error per antenna, $\phi$, required for an image of dynamic range $D$ is $\phi \approx N/\sqrt{2}D$, which is $\approx 1$~degree for the dynamic range of $>10^5$ required for the DSA-2000 (with $N = 2000$). This error incorporates residual phase errors between antennas, phase errors within the primary beam of each antenna (direction-dependent effects) and ionospheric phase errors across long baselines. Calibration will utilize pre-existing catalogs (e.g., NVSS and VLASS) and precisely measured antenna beam patterns, to solve for direction-independent gain terms, antenna pointing offsets and for reconstruction of the ionospheric phase screen above the array (e.g. Bhatnagar \& Cornwell 2017), utilizing a GPU-based realization of the Radio Interferometric Measurement Equation (RIME). Calibration/imaging via Bayesian inference techniques will also be explored during the development phase of the project. 

\item A wideband LNA and wide-band quad-ridge flared horn feed spanning 0.7--2\,GHz, coupled with modified dish optics is required to deliver a system temperature of 25\,K and aperture efficiency of $70\%$ across the DSA-2000 band. An average LNA temperature of 6\,K from 0.7--2\,GHz is required, which has been demonstrated within the DSA-110 band ($1.28 - 1.53$\,GHz) with a custom-designed LNA (Weinreb et al., in prep.) at room temperature. 

\item The correlator will be the largest yet built (a factor of ten larger than CHIME), and will incorporate a GPU-based flagging, direction-dependent calibration, gridding and imaging pipeline. Certain steps used in interferometric imaging have been demonstrated on a GPU, e.g. gridding (Van der Tol et al. 2018), but a full end-to-end pipeline has yet to be developed. 

\end{itemize}

\section{Organization, Partnerships, and Current Status}

The DSA-2000 is proposed to be designed, built and operated by a university-led consortium with US and international partners, together with the NANOGrav collaboration. A 3-year construction effort is proposed to take place following a 2.5-year design phase, during which the project office will also be established. This will draw on the project management heritage of the ongoing DSA-110 and NANOGrav projects.  The DSA-2000 will be proposed to the NSF Mid-scale Research Infrastructure-2 program with a view to first light in 2026. Additional construction funds beyond the \$70 million cap of the MSRI program will be via partner contribution. 

 \begin{figure}[!ht]
  \begin{center}
    \includegraphics[width=\textwidth]{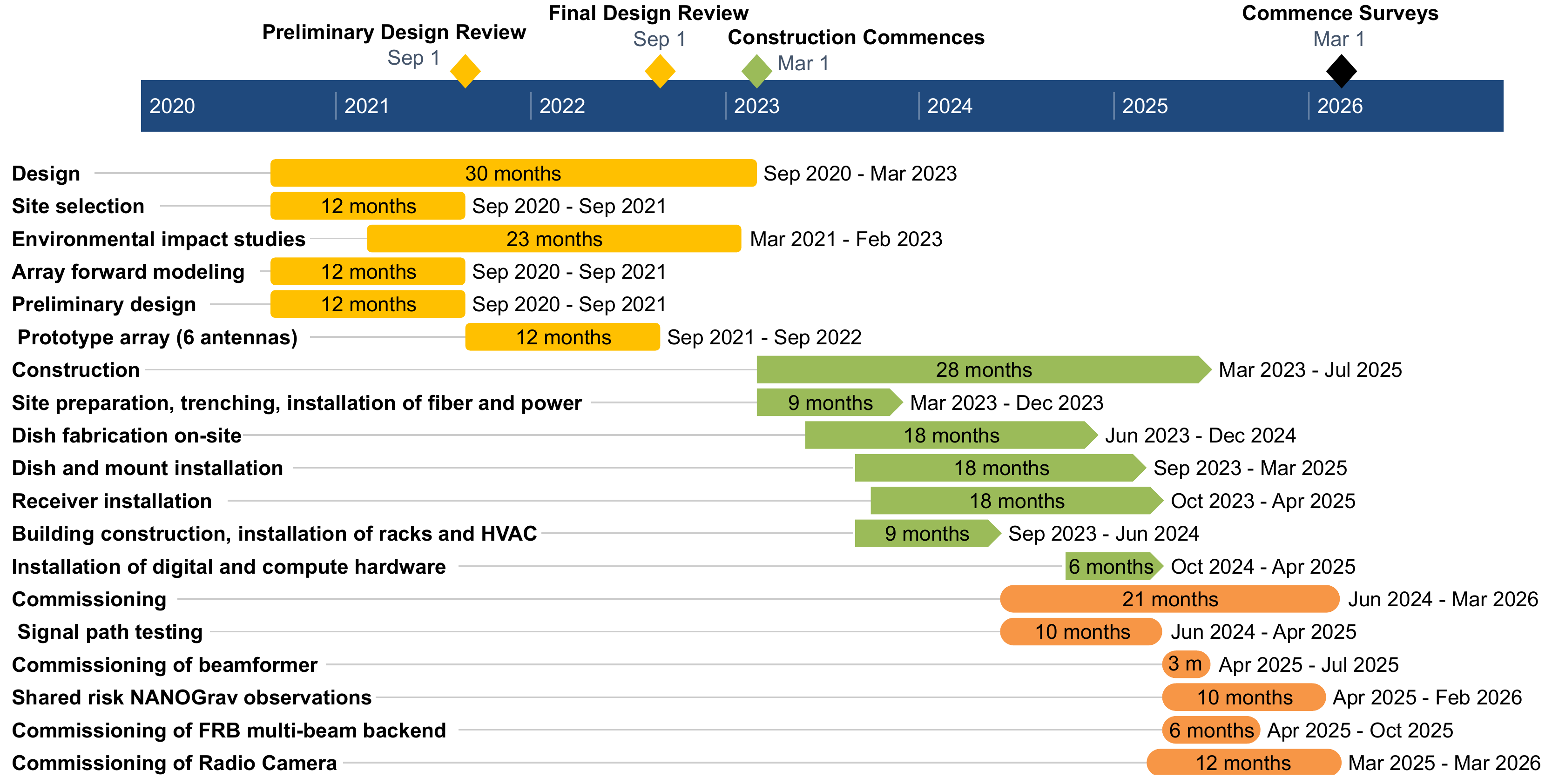}\\
    \textbf{Schedule for the DSA-2000 Project}
  \end{center}
  \label{fig:Gantt}
\end{figure}

\section{Cost Estimates}

The cost for design, construction and operation is summarized in Table 2. Cost per subsystem is inclusive of labor costs. The total cost for construction, including contingency, is \$96.25 million.

Architecture costs are inclusive of an Environmental Impact Study (EIS) and assume a pristine site; the latter incurs a large cost for installation of underground power cables for up to 10 km (\$6 million +50\% contingency).  Also included are the costs for a central processing building, including HVAC to dissipate up to 200\,kW, and the power and fiber cables trenched out to the location of each antenna. A contingency of 50\% is maintained for this subsystem to account for the uncertainty in costs relating to initial site development. 

Costing of the antennas package, including dish, mount, drive, receiver and monitor/control scales from the existing DSA-110 with aforementioned modifications. Similarly the digital hardware (including data storage and tape backups) scales from existing DSA-110 hardware and assumes 2019 pricing. Both subsystems are relatively low cost risk and assume 25\% contingency.
DSA-2000 data management cost drivers include: all-sky spectral cubes for an estimated 5-year data volume of 500\,TB, mirrored in two geographically distinct locations; a public image service and associated multi-band / multi-epoch catalog serving single combined all-sky epoch images, sized at 3TB; Pulsar Timing Array data products from 200 pulsars observed monthly, with raw and calibrated profiles totaling approximately 150\,TB a year for a five year total of 750\,TB mirrored in two locations, publicly available after 18-month proprietary period, with times-of-arrival and ephemerides publicly available at time of publication.

Total cost for computer hardware in two geographically distinct locations is estimated at \$3-400,000. Programmer cost and systems administration cost for production of the necessary services is estimated at 3 university or facility-based FTEs (potentially split over several fractional FTEs), with an addition FTE for technical oversight/senior programmer for two years, estimated at \$1.8 million. The total construction cost is estimated at \$2.2-2.3 million. 

Operations costs include technicians, mechanical engineers, machinists and RF engineers for the maintenance of 2000 dishes, as well as ground, building and road maintenance. Network/compute administrative costs, and software and digital engineer support is costed for the DSA-2000 backend. Power costs include 70\,kW for the correlator/beamformer FPGAs, 50\,kW for the GPU servers used for gridding/imaging and 30\,kW for storage/networking hardware. Each dish is assumed to require 500\,W for continuous tracking and slewing. Thus the total power requirement is 1.15\,MW. Assuming \$0.1 per kW/hr, results in a cost estimate of \$1 million/year. Annual operations cost for the public archive is figured at 1.5 FTEs for programmers and systems administration plus one additional FTE for technical oversight/senior programmer, and hosting costs for the computer hardware, estimated at \$600k. Cloud services could comprise at least one of the mirrored copies and provide hosting for the service components. 25\% contingency is assumed for operations costing. 

\begin{table}[ht]
\caption{\small DSA-2000 costs.}
\normalsize
\begin{center}
 \begin{tabular}{||c | c | c | c||} 
 \hline
& Estimated Cost & Contingency & Total \\
\hline
\hline
\textbf{Design/Development} & \textbf{\$3.8 million}  & \textbf{25\%} & \textbf{\$4.75 million} \\ \hline
Antennas, Mounts and Drives & \$28 million  & 25\% & \$35 million \\
Receivers & \$4 million  & 25\% & \$5 million \\
Infrastructure and Buildings & \$15.2 million  & 50\% & \$22.8 million \\
Digital Backend ($+$ imaging, storage) & \$25 million  & 25\% & \$30 million \\
Public Archive  & \$2.3 million  & 50\% & \$3.45 million \\
\textbf{Total Construction}  & \textbf{\$74.5 million}  & \textbf{30\%} & \textbf{\$96.25 million} \\ 
\hline
\textbf{Operations/yr}  & \textbf{\$5.3 million}  & \textbf{25\%} & \textbf{\$6.6 million} \\ \hline

\end{tabular}
\end{center}
\label{tab:t2}
\end{table}




\clearpage 

\section{References}

\begin{enumerate}
    
    \item Becker, Robert H.; White, Richard L.; Helfand, David J. 1995, ApJ, 450, 559 -- `The FIRST Survey: Faint Images of the Radio Sky at Twenty Centimeters'
    \item Bhatnagar, S.; Cornwell, T. J. 2017, AJ, 154, 197B -- `The Pointing Self-calibration Algorithm for Aperture Synthesis Radio Telescopes'
    \item Condon, J. J.; Cotton, W. D.; Greisen, E. W. et al. 1998, AJ, 115, 1693 -- `The NRAO VLA Sky Survey'
    \item Condon, J. J.; Cotton, W. D.; Fomalont, E. B. et al. 2012, ApJ, 758, 23 -- `Resolving the Radio Source Background: Deeper Understanding through Confusion'
    \item Cotton, W. D.; Condon, J. J.; Kellermann, K. I. et al. 2018, ApJ, 856, 67 -- `The Angular Size Distribution of $\mu$Jy Radio Sources'
    \item Decarli, R.; Walter, F.; Venemans, B. P. et al. 2017, Nature, 545, 457 -- `Rapidly star-forming galaxies adjacent to quasars at redshifts exceeding 6'
    \item Hallinan, G.; Corsi, A.; Mooley, K. P. et al. 2017, Science, 358, 1579 -- `A radio counterpart to a neutron star merger'
    \item Hotokezaka, Kenta; Nakar, Ehud; Gottlieb, Ore et al. 2019, arXiv:1806.10596 -- `A Hubble constant measurement from superluminal motion of the jet in GW170817'
    \item Kocz, J.; Ravi, V.; Catha, M. et al. 2019, arXiv:1906.08699 -- `DSA-10: A Prototype Array for Localizing Fast Radio Bursts'
    \item Law, Casey; Gaensler, Bryan; Metzger, Brian, et al. 2018, ApJL, 866, L22 -- `Discovery of the Luminous, Decades-long, Extragalactic Radio Transient FIRST J141918.9$+$394036 ' 
    \item Magliocchetti, M.; Popesso, P.; Brusa, M. et al. 2018, MNRAS, 473, 2493 -- `A census of radio-selected AGNs on the COSMOS field and of their FIR properties'
    \item Mezcua, M.; Suh, H.; Civano, F. 2019, MNRAS accepted, arXiv:1906.10713 -- `Radio jets from AGN in dwarf galaxies in the COSMOS survey: mechanical feedback out to redshift $\sim3.4$'
    \item Mooley, K. P.; Hallinan, G.; Bourke, S. et al. 2016, ApJ, 818, 105 -- `The Caltech-NRAO Stripe 82 Survey (CNSS). I. The Pilot Radio Transient Survey In 50 deg2'
    \item Mooley, K. P.; Deller, A. T.; Gottlieb, O. et al. 2018, Nature, 561, 355 -- `Superluminal motion of a relativistic jet in the neutron-star merger GW170817'
    \item Nakar, Ehud; Piran, Tsvi 2011, Nature, 478, 82 -- `Detectable radio flares following gravitational waves from mergers of binary neutron stars'
    \item Ravi, V.; Catha, M.; D'Addario, L. et al. 2019, \href{https://doi.org/10.1038/s41586-019-1389-7}{Nature Accelerated Article Preview}, \\ arXiv:1907.01542 -- `A fast radio burst localised to a massive galaxy'
    \item van der Tol, Sebastiaan; Veenboer, Bram; Offringa, Andre R. 2018, A\&A, 616, A27 -- `Image Domain Gridding: a fast method for convolutional resampling of visibilities'
    \item Wilman, R. J.; Miller, L.; Jarvis, M. J. et al. 2008, MNRAS, 388, 1335 -- `A semi-empirical simulation of the extragalactic radio continuum sky for next generation radio telescopes'

\end{enumerate}

}
\end{document}